\documentclass[twocolumn]{aastex61hack}

\newcommand{\rii}{\emph{r}-II}
\newcommand{\ri}{\emph{r}-I}
\newcommand{\rp}{\emph{r}-process}
\newcommand{\ThDy}{$\log\epsilon({\rm Th/Dy})$}
\newcommand{\ZrDy}{$\log\epsilon({\rm Zr/Dy})$}

\usepackage{enumerate}
\usepackage{graphicx}
\usepackage{amsmath}
\usepackage{natbib}
\usepackage{float}
\usepackage{hyperref}

\begin{document}

\title{Actinide-rich and Actinide-poor \emph{r}-Process Enhanced Metal-Poor Stars\\ do not Require Separate \emph{r}-Process Progenitors}

\author{Erika M.\ Holmbeck}
\affiliation{Department of Physics, University of Notre Dame, Notre Dame, IN 46556, USA}
\affiliation{JINA Center for the Evolution of the Elements, USA}

\author{Anna Frebel}
\affiliation{Department of Physics and Kavli Institute for Astrophysics and Space Research, Massachusetts Institute of Technology, Cambridge, MA 02139, USA}
\affiliation{JINA Center for the Evolution of the Elements, USA}

\author{G.\ C.\ McLaughlin}
\affiliation{Department of Physics, North Carolina State University, Raleigh, NC 27695, USA}
\affiliation{JINA Center for the Evolution of the Elements, USA}

\author{Matthew R.\ Mumpower}
\affiliation{Theoretical Division, Los Alamos National Laboratory, Los Alamos, NM, 87545, USA}
\affiliation{Center for Theoretical Astrophysics, Los Alamos National Laboratory, Los Alamos, NM, 87545, USA}\affiliation{JINA Center for the Evolution of the Elements, USA}

\author{Trevor M.\ Sprouse}
\affiliation{Department of Physics, University of Notre Dame, Notre Dame, IN 46556, USA}

\author{Rebecca Surman}
\affiliation{Department of Physics, University of Notre Dame, Notre Dame, IN 46556, USA}
\affiliation{JINA Center for the Evolution of the Elements, USA}

\correspondingauthor{Erika M.\ Holmbeck}
\email{eholmbec@nd.edu}

% =====================================================
% Abstract
% =====================================================

\begin{abstract}
The astrophysical production site of the heaviest elements in the universe remains a mystery.
Incorporating heavy element signatures of metal-poor, \rp\ enhanced stars into theoretical studies of \rp\ production can offer crucial constraints on the origin of heavy elements.
In this study, we introduce and apply the ``Actinide-Dilution with Matching" model to a variety of stellar groups ranging from actinide-deficient to actinide-enhanced to empirically characterize \rp\ ejecta mass as a function of electron fraction.
We find that actinide-boost stars do not indicate the need for a unique and separate \rp\ progenitor.
Rather, small variations of neutron richness within the same type of \rp\ event can account for all observed levels of actinide enhancements.
The very low-$Y_e$, fission-cycling ejecta of an \rp\ event need only constitute 10--30\% of the total ejecta mass to accommodate most actinide abundances of metal-poor stars.
We find that our empirical $Y_e$ distributions of ejecta are similar to those inferred from studies of GW170817 mass ejecta ratios, which is consistent with neutron-star mergers being a source of the heavy elements in metal-poor, \rp\ enhanced stars.
\end{abstract}


%\received{}
%\revised{}
%\accepted{}
%\published{}
\submitjournal{\apj}

% =====================================================
\section{Introduction}

The rapid-neutron capture (``\emph{r-}'') process is thought to be a main mechanism to synthesize elements heavier than iron and the only mechanism capable of producing the actinide elements, such as thorium and uranium.
Astrophysically, possible sites of the \rp\ remain unconfirmed; core-collapse supernovae (CCSNe) and neutron-star mergers (NSMs) are the long-favored candidates.
CCSNe were thought to be natural sites for robust \rp\ production since \citet{b2fh}.
Although several studies have shown that current models of CCSNe cannot reproduce the heavy/main elemental \rp\ pattern of the Solar System, they may still be responsible for the light \rp\ elements \citep{Thielemann+11,arcones2013}.
The recent NSM event GW170817 \citep{abbott2017} and corresponding electromagnetic afterglow AT\,2017gfo \citep{kilpatrick2017,cowperthwaite2017,drout2017,shappee2017} now lends additional observational support for NSMs as robust producers of lanthanide material. 
Earlier observational evidence in support of NSM was the discovery of the ``\rp\ dwarf galaxy" Reticulum~II (``Ret~II").
In this ultra-faint dwarf galaxy (UFD), high-resolution spectroscopic studies \citep{ji2016,roederer2016} identified multiple low-metallicity stars with extreme \rp\ enhancement. Significant \rp\ enrichment in such a small system calls for an event that ejected large amounts of \rp\ material, which \citet{ji2016b} argue could not be from standard SNe, but could be explained with an NSM.

Simulations suggest that one NSM event houses several environments capable of undergoing an \rp.
Of particular importance are the neutron-rich, low-entropy dynamical (or tidal) ejecta, which escape at high velocities \citep{lattimer1974,Meyer89,Freiburghaus+99}.
On slightly longer timescales is the accretion disk wind, which is estimated to have slightly lower neutron-richness and higher entropy than the tidal ejecta \citep{surman2008,metzger2008,perego2014}.
Neutrino flavor transformation has the potential to make the wind significantly more neutron rich than currently predicted by simulation \citep{malkus2016}.
An accretion disk wind that may facilitate an \rp\ is not limited to just NSM environments.
Recently, the accretion disk around collapsars---the core-collapse of a massive rotating star---has seen a resurgence as a possible site of robust \rp\ element production \citep{pruet2004,surman2004,siegel2018}.
Other \rp\ sites have also been proposed, such as magneto-rotational instability-driven SNe (\citealt{cameron2003,winteler2012,nishimura2015}, but see also \citealt{mosta2018}), and dark matter-induced neutron star implosions \citep{bramante2016,fuller2017}.

A well-established method for obtaining empirical evidence on \rp\ sites is through observations of metal-poor stars in the Milky Way halo that are strongly enriched in \rp\ elements.
The ``\rii" stars (defined as ${\rm[Eu/Fe]}>+1.0$ and $[{\rm Ba/Eu}]<0$) display a strong relative enhancement of \rp\ elements in their photospheres compared to their iron content \citep{barklem2005,beers2005}.
About 3--5\% of stars in the Milky Way halo with ${\rm [Fe/H]}\lesssim-2$ are classified as \rii, totaling about 30 \rii\ stars identified as of 2015 \citep[from data in][``JINAbase"\footnote{\url{https://github.com/abduabohalima/JINAbase}}]{abohalima2017}.
Outside the Milky Way, about ten UFDs had been studied for \rp\ enrichment as of 2016.
At that time, only one---Ret~II---was found to have \rii\ stars, with seven (of nine observed) stars identified as \rii\ members \citep{ji2016,roederer2016}.
Given the hierarchical merger origin of the Milky Way \citep{searle1978,schluafman2009,tumlinson2010}, metal-poor halo stars likely formed in small early galaxies such as analogs of the surviving UFDs.
Accordingly, such strong \rp\ enhancement in halo stars suggests that \rp\ events occurring in these galaxies, such as NSMs that eject large amounts of \rp\ material, should overall be favored as early \rp\ production sites.

Besides the \rii\ stars, there are also the moderately enhanced ``\emph{r}-I" metal-poor stars ($+0.3<{\rm[Eu/Fe]}\leq +1.0$ and $[{\rm Ba/Eu}]<0$).
These stars possibly formed in somewhat larger dwarf galaxies, such as Tucana~III \citep{hansen2017}, in which the yields of any prolific \rp\ event would be diluted more than in the case of the formation of \rii\ stars in smaller systems.
The range of both metallicites and level of \rp\ enrichment at which the \ri\ and \rii\ stars are found suggests that NSMs alone could not account for all the \rp\ material in the Galaxy.
As \citet{cote2018b} argue, it is likely that a separate site (or sites) could have contributed \rp\ material at early times in the universe.

The \ri\ and \rii\ stars show striking similarities in their main \rp\ patterns among the lanthanide elements ($_{57}$La through $_{71}$Lu).
However, some variation exists in the actinide elements, Th and U, with about 30\% showing an enhancement of Th relative to the lanthanides \citep{mashonkina2014}, dubbed the ``actinide-boost" stars.
There is also a wider variation of the elemental abundances that follow the first \rp\ peak---Sr, Y, and Zr---with respect to their scaled main \rp\ abundances \citep{siqueira2014,ji2016b}.
Due to these variations, it is thought that Sr--Y--Zr may originate from a different \rp\ environment than what produces the lanthanides and actinides, such as the limited-\rp, which would primarily synthesize $Z<56$ elements \citep{travaglio2004,hansen2012,arcones2013,wanajo2013,frebel2018}.
Similarly, the actinide variation may indicate a separate \rp\ progenitor object or site that is responsible for the existence of actinide-boost stars \citep{schatz2002}.

Alternatively, it may be possible that the variations in the actinides and limited-\emph{r} elements in the \ri\ and \rii\ stars can be fully accounted for by variations of astrophysical conditions (e.g., the electron fraction, $Y_e$) within the same \rp\ source (i.e., type of site). 
In this work, we identify key elemental abundance measurements of metal-poor \rp\ enhanced stars to give insight into the progenitor \rp\ events that gave rise to the observed abundance variations.
These key measurements are used in concert with our theoretical ``Actinide-Dilution with Matching" model to ascertain whether the existence of actinide-boost stars suggests one distinct \rp\ site or if the range of (relative) actinide element abundances can be plausibly explained by a continuum of conditions within the same type of source.
With this analysis in hand, we are able to weigh in on the implications of the observations of limited-\emph{r} and actinide elements, and further, to use observations of low-metallicity stars to provide a consistency check on the amount of lanthanide-rich material inferred from recent ``kilonova" observations.

In Section~\ref{sec:observations}, we discuss \rp\ patterns of metal-poor stars and quantify distinct differences in their scaled abundances that could reflect different \rp\ sites or conditions among the earliest \rp\ events.
In Section~\ref{sec:adm}, we introduce and detail our Actinide-Dilution with Matching model.
Next, we apply this model to different groups of \rp\ enhanced stars that were likely enriched by just one event, and we present these results in Section~\ref{sec:results}.
In Section~\ref{sec:variations}, we investigate variations on the astrophysical and nuclear inputs that could affect our model results.
Finally in Section~\ref{sec:gw710817}, we compare our empirical $Y_e$ distributions of mass ejecta to that of the GW170817 associated kilonova to test if our results align with these recent observations.

% =====================================================
\section{Observations of Metal-Poor Stars}
\label{sec:observations}

In this section, we discuss observations of metal-poor stars in the context of actinide and limited-\emph{r} production.
To study the full range of the elemental \rp\ pattern at early times, we choose Zr, Dy, and Th as representative of the limited-\emph{r} process, main \rp, and actinides, respectively.
Although $_{38}$Sr and $_{63}$Eu are traditionally used to quantify the limited-\emph{r} and main \rp\ contributions, we instead use $_{40}$Zr and $_{66}$Dy to probe these two regions.
More and unsaturated absorption lines of \ion{Zr}{2} are available over the few of \ion{Sr}{2} from which to derive an abundance, leading to Zr abundances with higher precision.
In addition, \ion{Sr}{2} suffers larger systematic abundance corrections from assuming local thermodynamic equilibrium (LTE) over non-LTE, while the \ion{Zr}{2} corrections are lower and the abundances more robust under LTE \citep{andrievsky2011}.

In the lanthanide region, the production of Eu by the \rp\ may be sensitive to fission yields, especially to broad and asymmetric fission distributions that place material above the second \rp\ peak \citep[e.g.,][]{kodama1975,eichler2015,Cote+18,vassh2018}.
Moreover, the fission fragment distributions of nuclei that may participate in the \rp\ at high nuclear masses are far from known.
To avoid fission-dependent results, we use Dy instead of Eu.
At a slightly higher mass, Dy is nearly insensitive to the direct effects of fission fragment distributions.

% #========================================
\subsection{Milky Way \emph{r}-Process Enhanced Stars}
\label{sec:mwstars}

\begin{figure}[t]
	\centering
    \includegraphics[width=0.9\columnwidth]{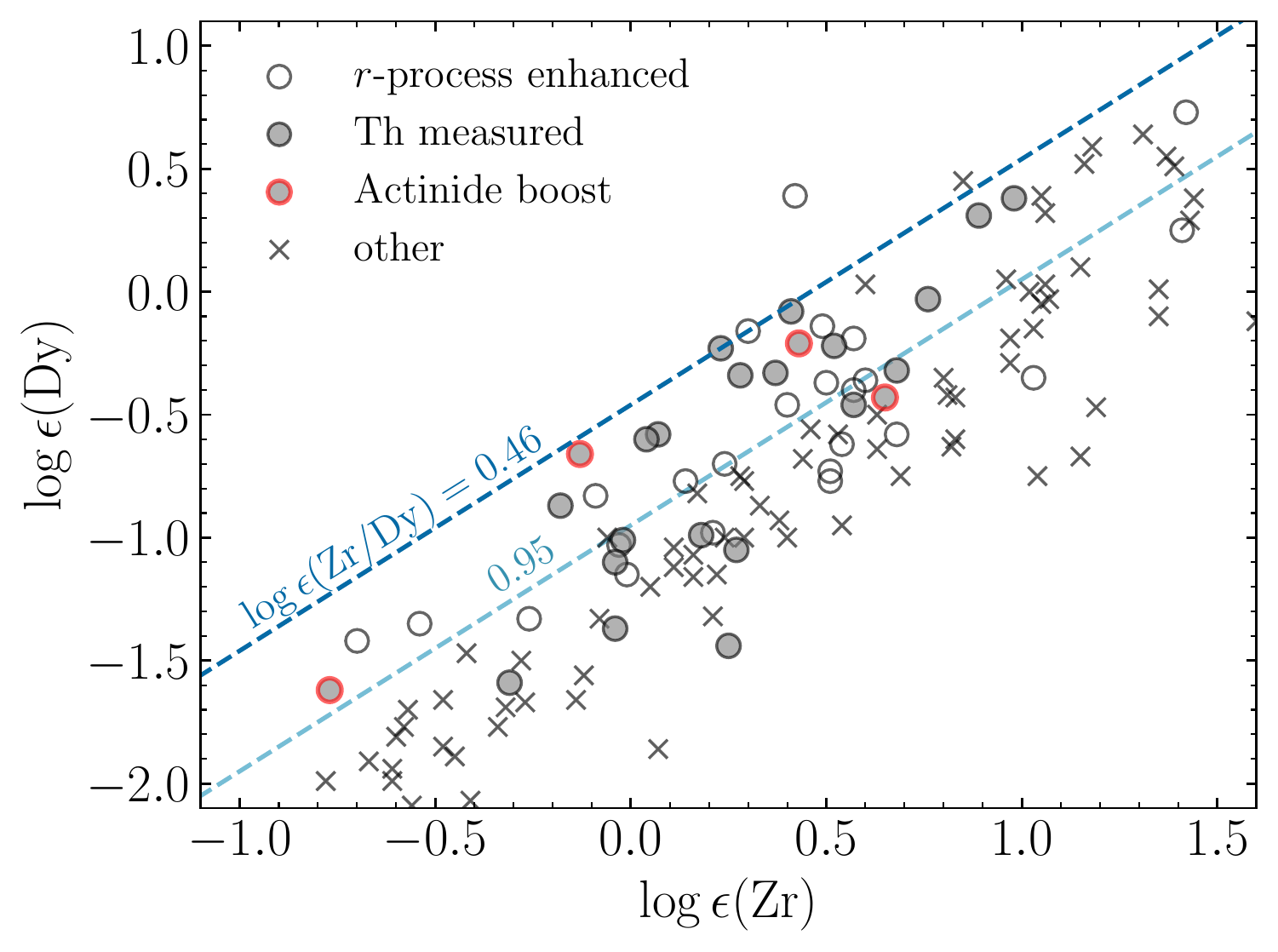}
    \includegraphics[width=0.9\columnwidth]{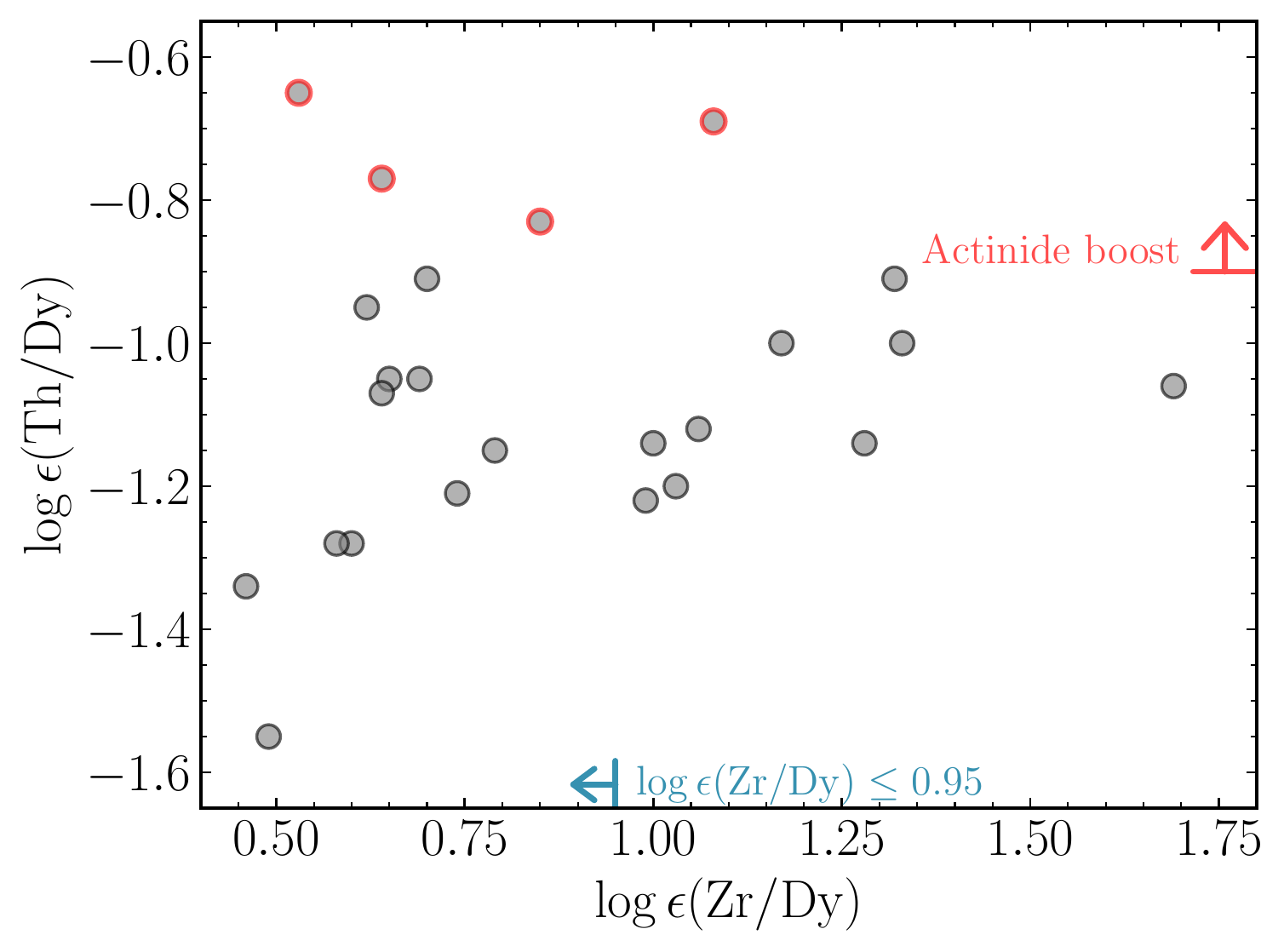}
    \caption{Top: Dy versus Zr abundances for metal-poor Milky Way stars with an \rp\ signature (circles), a measurement of Th (filled circles), an actinide-boost signature (red outlines), or other enrichment (e.g., \emph{s}-process, \emph{i}-process, and/or carbon enhancement; crosses). The dashed lines are constant values of \ZrDy\ of 0.46 and 0.95. Bottom: scatter of \ThDy\ as a function of \ZrDy\ for those stars with a measurement of Th. Data were selected from \citet{abohalima2017,placco2017,ji2018,sakari2018,holmbeck2018}.\label{fig:scatter}}
\end{figure}

We first consider all metal-poor Milky Way stars that have both Zr and Dy abundance measurements included in JINAbase and individual additions from the recent discoveries in \citet{placco2017}, \citet{ji2018}, \cite{sakari2018}, and \citet{holmbeck2018}.
This data set is displayed in the top panel of Figure~\ref{fig:scatter}.
The absence of stars with both low Zr and high Dy abundances (i.e., upper-left of the top panel of Figure~\ref{fig:scatter}) may suggest that some minimum Zr is made in the same event that created Dy.
This trend was also noted in \citet{roederer2013} who analogously used Sr and Ba abundances.

The light and dark blue dashed lines in Figure~\ref{fig:scatter} indicate $\mbox{\ZrDy}= 0.95$ and $\mbox{\ZrDy}= 0.46$, respectively, solely for reference and guidance on the abundance trend.
Stars with $\mbox{\ZrDy}> 0.95$ are mostly those with no \rp\ enhancement (i.e., $[{\rm Eu/Fe}]\leq +0.3$) and/or with enhancement in other elements, such as carbon and \emph{s}-process elements (e.g., $[{\rm Ba/Eu}] \geq 0$).
The line at $\mbox{\ZrDy}=0.46$ reflects that of the scaled, average \ZrDy\ abundance for \rp\ stars in Ret~II.
We note that all stars with a Th measurement have a \ZrDy\ abundance of at least this value.
The bottom panel of Figure~\ref{fig:scatter} shows the subset of stars from the top panel that, in addition, have a Th measurement.
The wide range of \ThDy\ abundances is entirely represented by stars with $\mbox{\ZrDy}\leq 0.95$.
At higher values of \ZrDy, the \ThDy\ appears to converge towards a constant value of $\mbox{\ThDy}\approx -1.0$.

Most of the confirmed \rp\ enhanced stars lie in the range $0.46\leq \mbox{\ZrDy}\leq 0.95$. These stars also show the broadest range of \ThDy. For this work, we posit that these \rp\ stars display a pure \rp\ signature that has come from just one event. 
For \rp\ stars with $\mbox{\ZrDy}> 0.95$, while it is possible that their \rp\ signatures may have also come from a single event, it is also possible that their \rp\ material has been diluted or altered by additional types of nucleosynthesis (i.e., other than a main \rp) or strong contributions from limited-\emph{r} process events.
Therefore, to study the widest range of actinide production by a single \rp\ site, we focus on \rp\ stars with $\mbox{\ZrDy}\leq 0.95$.

% #======================================
\subsection{Kinematically Linked Groups of \emph{r}-Process Enhanced Stars}
\label{sec:kinematic_groups}

Given the presumed accretion of stars that now reside in the Milky Way's halo, the \rp\ enhanced halo stars have essentially unknown origins. Specifically, it has been suggested that the \rp\ enhanced halo stars originated in dwarf galaxies that were eventually accreted by the Milky Way as part of its hierarchical growth.
If a prolific \rp\ event enriched the original, low-mass host galaxy, such as that in Ret II, the imprints on these stars offer a window into the element production by (presumably) single \rp\ events.

\citet{roederer2018} recently found kinematic grouping among spatially unrelated \rp\ enhanced halo stars.
These kinematic groups are further evidence that \rp\ enhanced halo stars were once members of satellite galaxies which became accreted by the Milky Way.
The progenitor dwarf galaxies of these kinematic groups could resemble Ret~II, where all stars belonging to each of these progenitor systems would have formed from gas enriched by single, respective \rp\ events.
Therefore, we assume that the elemental abundances of stars in the kinematic groups now reflect the range of element production by single events.
Abundance pattern differences among members of each groups could then point to different astrophysical \rp\ conditions within the same type of event or even entirely different \rp\ sources.
In this regard, the seven \rii\ stars in Ret~II can be treated as an additional such group as it is highly likely that only one \rp\ event took place prior to their formation.
Hence, stellar abundance variations within these groups could provide insight into the range of \rp\ element production by a single event.

In the following, we expand on the principal idea of assigning groups of \rp\ stars.
Specifically, we focus on elemental abundance variations between these groups in the actinide and limited-\emph{r} elements.
Here, we define ``actinide-deficient" as $\mbox{\ThDy} < -1.20$, ``actinide-normal" as $-1.20 \leq \mbox{\ThDy} \leq -0.90$, and ``actinide-boost" as $\mbox{\ThDy} > -0.90$.

\emph{Ret~II} --- Although the scaled, heavy-element (between Ba and the third peak) abundance patterns of seven Ret~II stars closely resemble those of \rii\ halo stars, the only Ret~II member for which a Th measurement is available \citep[DES~J033523$-$540407;][]{ji2018} displays a strikingly low actinide abundance compared to its lanthanides ($[{\rm Th/Eu}]=-0.34$).
The seven \rii\ stars of Ret~II might reflect an event with low actinide production, or possibly one with a significant range.
Without a complete set of Th abundances for each of the \rii\ stars in Ret~II, we assume, for simplicity, that this low actinide level reflects low actinide production in the \rp\ event that enriched the Ret~II gas.
Thus we assume Ret~II has $\mbox{\ThDy} = -1.49$.

\emph{Group F} --- The kinematic ``Group~F" in \citet{roederer2018} consists of three stars: {CS~29529-054} \citep{roederer2014,roederer2014b}, HE~2224+0143 \citep{barklem2005,ren2012}, and HD115444 \citep{westin2000}, the latter two of which have ``normal" actinide abundances: $[{\rm Th/Eu}]=0.05$ and $[{\rm Th/Eu}]=-0.21$, respectively, and $\mbox{\ThDy} = -1.19$ on average.

\emph{J0954+5246} --- Just a single star, but representing extreme levels of actinide production by an \rp. 2MASS~J09544277+5246414 \citep[``J0954+5246";][]{holmbeck2018} is currently the most actinide-enhanced \rii\ star known, with $[{\rm Th/Eu}]=0.38$ and $\mbox{\ThDy} = -0.65$.

We treat these three levels of relative actinide enhancement as three distinct ``groups" and assume that each group's members formed from gas enriched by a individual \rp\ event.
Together, the stellar abundances of the stars in Ret~II, %($N=7$) 
Group~F,
%($N=3$)
and J0954+5246 
%($N=1$)
reflect a range of actinide enhancement, which may indicate either separate \rp\ actinide sources or a variation within one type of \rp\ source.

\begin{figure}[t]
	\centering
    \includegraphics[width=\columnwidth]{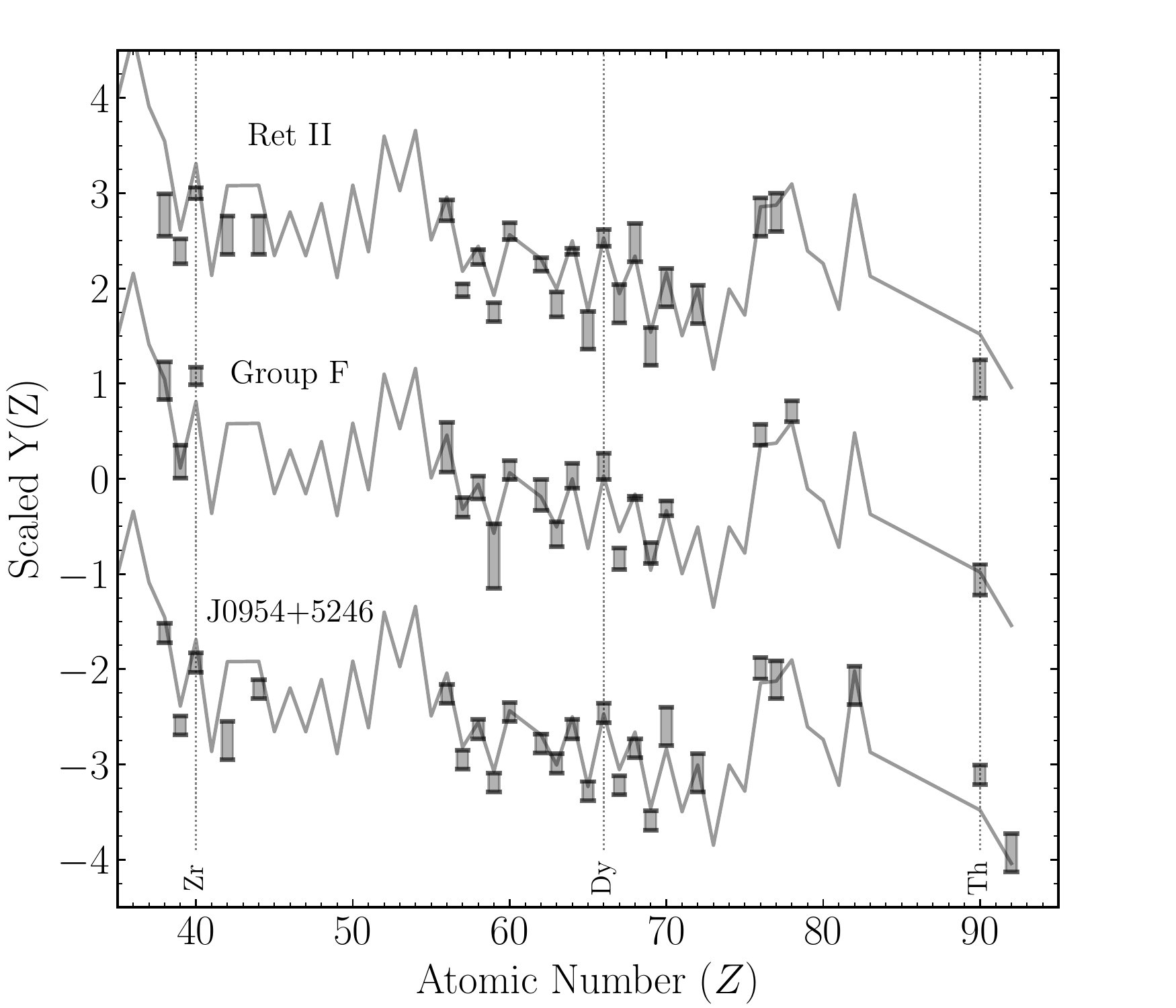}
    \caption{Abundance ranges (arbitrary scaling) of the three stellar groups considered in this work. Solid gray lines show the scaled Solar \rp\ pattern.\label{fig:patterns}}
\end{figure}

Between the three groups, the abundances of the limited-\emph{r} elements (Sr, Y, and Zr) also vary with respect to the lanthanide abundances. Whereas it has been suggested that these light neutron-capture elements may originate from a separate \rp\ site, we assume in this analysis that for \rp\ enhanced stars with $\mbox{\ZrDy} \leq 0.95$, these elements come from the same event that also synthesized the actinides.
Thus, within each group, we consider the relative variations among the limited-\emph{r} elements as well as the actinides as intrinsic to the progenitor \rp\ event.

For this study, we combine the abundances of stars within Ret~II and Group~F by scaling the individual abundance patterns to the respective average residual obtained from comparison with the Solar \rp\ pattern between $_{56}$Ba and $_{71}$Lu.
After scaling the Solar pattern such that the average deviation of the stellar pattern from Solar pattern between Ba to Lu is minimized,
we find the range of scaled abundances derived for each element over all stars in Ret~II and Group~F separately.
For J0954+5246, and in the cases where an element was only measured in one star in the group (e.g., Th in Ret~II), we use the reported uncertainty in its derived abundance as representative of the ``range" for the group.
These ranges/uncertainty bands are displayed in Figure~\ref{fig:patterns} for the three enrichment cases.
In Section~\ref{sec:adm}, we adopt these scaled and combined abundance values as model input, in order to reconstruct possible distributions of \rp\ material ejected by each of the putative progenitor \rp\ events.

% =====================================================
\section{The Actinide-Dilution with Matching Model}
\label{sec:adm}

The electron fraction ($Y_e$) is a major factor governing the ultimate extent of element production by an \rp\ event.
Variations of how \rp\ ejecta mass is distributed in $Y_e$ may explain the abundance variations within and between stellar groups of \rp\ enhanced stars, as those described in Section~\ref{sec:observations}.
In this section, we build empirical \rp\ ejecta distributions as a function of $Y_e$ by employing a Monte-Carlo method as an extension to the Actinide-Dilution (``AD") model introduced in \citet{holmbeck2018b}, which we call the Actinide-Dilution with Matching (``ADM") model.
To constrain the model by matching results to stellar abundances, we use three particular regions of the observed \rp\ elemental abundance patterns: the limited-\emph{r} group, the lanthanides, and the actinides, represented by Zr, Dy, and Th, respectively.
These abundance constraints and their allowed tolerances for the ADM model results are listed in Table~\ref{tab:ratios} when using the three groups described in Section~\ref{sec:kinematic_groups}.

\begin{table}[t]
	\centering
    \caption{Abundance ratio matching conditions used by the ADM method for each stellar group considered in this work.\label{tab:ratios}}
    \footnotesize
	\begin{tabular}{l c c c c} \hline\hline
    Group & $N$ & $\log\epsilon({\rm Zr/Dy})$ & $\log\epsilon({\rm Th/Dy})$ & $\log\epsilon({\rm U/Th})$ \\ \hline
    Ret II & 7 & $0.46 \pm 0.20$ & $-1.49\pm 0.30$ & $-0.25 \pm 0.10$ \\
    Group F & 3 & $0.95 \pm 0.20$ & $-1.19 \pm 0.30$ & $-0.25 \pm 0.10$ \\
    J0954+5246 & 1 & $0.53 \pm 0.20$ & $-0.65 \pm 0.30$ & $-0.25 \pm 0.10$ \\
    \hline
    \end{tabular}
\end{table}

Since Th could only be measured in one or two stars per group, the allowed abundance ratios listed in Table~\ref{tab:ratios} come from a single star with the assumption that all other stars within the group have $\log\epsilon({\rm Th/ Dy})$ ratios lying with a broad 0.3~dex of that single measurement.
Furthermore, we add 0.2 dex to the adopted \ThDy\ matching-constraint listed in Table~\ref{tab:ratios}.
This addition accounts for radioactive Th decay over roughly 10 Gyr from the final abundances of our \rp\ calculations to the present.

Of the three groups in Figure~\ref{fig:patterns}, only one star has a reliable uranium measurement, which is unsurprising given that overall, fewer than ten \rp\ enhanced stars have a reliable detection of uranium.
For stars with both Th and U measurements available, studies applying radioactive decay dating have shown the U/Th production ratio to be roughly constant, $\log\epsilon({\rm U/Th})\approx -0.25$, even for the actinide-boost stars which show absolute enhancement in these elements \citep[e.g.,][]{cowan1999,schatz2002,wanajo2002,farouqi2010}.
Hence, for this analysis, we assume that the \rp\ material in all stars with Th was produced with the same U/Th ratio, and supply this ratio as an additional constraint to the ADM model.
The \emph{production} ratio rather than the observed ratio is used since Th and U are radioactive, and their abundances change over time.

After establishing the observational constraints, we first ran several \rp\ simulations using a medium-entropy parameterized trajectory (evolution of an ejecta mass element, here with initial entropy $s/k\approx 40$ and dynamical timescale $\tau_{\rm dyn}=20~{\rm ms}$) as in \citet{zhu+2018}.
This trajectory is consistent with an accretion disk wind around a proto-neutron star \citep[e.g., a collapsar or NSM remnant;][]{surman2004}.
We vary the $Y_e$ as in \citet{holmbeck2018b} to allow for multiple levels of neutron-richness within the same environment, changing the initial $Y_e$ from 0.005 to 0.450 in equal steps of 0.005.
The \rp\ calculations are run using the nuclear network code Portable Routines for Integrated nucleoSynthesis Modeling \citep[PRISM;][]{mumpower2017,Cote+18,mumpower2018,vassh2018}.
Reaction and decay rates relevant to the \rp\ are constructed as self-consistently as possible. 
Starting with nuclear masses from the Finite Range Droplet Model \citep[FRDM2012;][]{moller2012,moller2016}, we adopt the neutron-capture and neutron-induced fission rates calculated self-consistently with FRDM2012 masses using the Los Alamos National Laboratory statistical Hauser-Feshbach code \citep{Kawano+16}.
The QRPA+HF framework \citep{mumpower2016} is used to calculate the relative probabilities of $\beta$-decay, $\beta$-delayed fission, and $\beta$-delayed neutron emission for each nucleus, using \citet{Moller+18} $\beta$-decay strength functions.
Fission barrier heights from \citet{moller2015} are used to calculate fission rates, employing the \citet{Zagrebaev+11} relation for the spontaneous fission channel and adopting symmetric fission fragment distributions for all fission channels.

\begin{figure}[t]
	\centering
    \includegraphics[width=0.9\columnwidth]{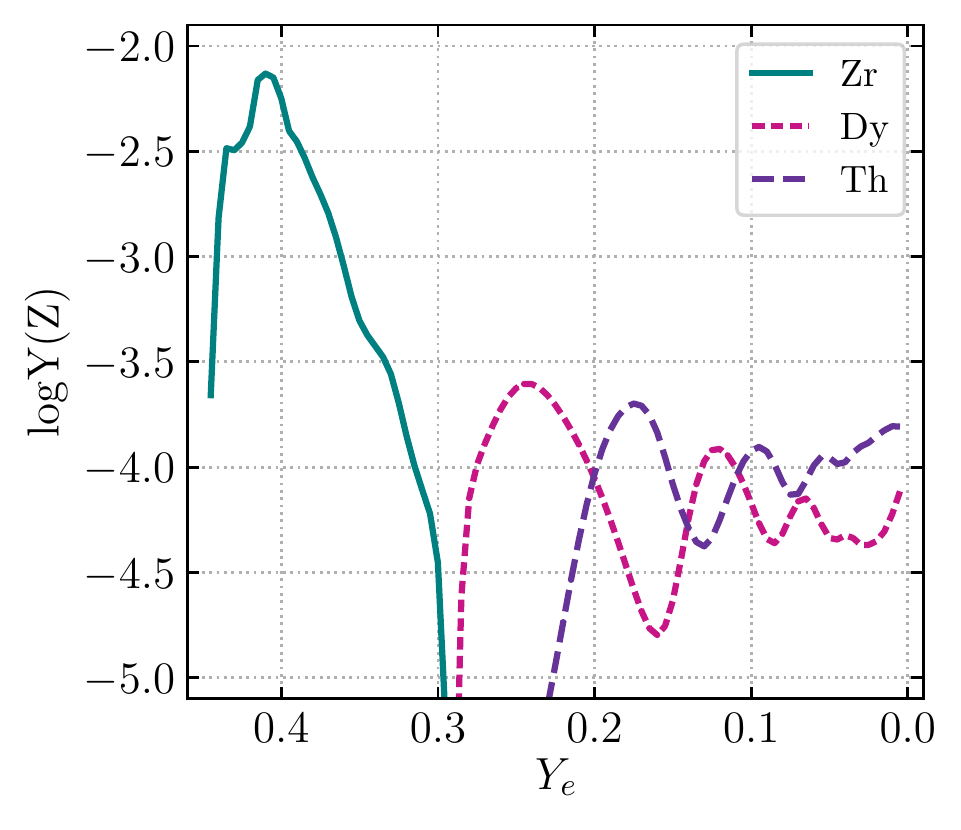}
    \caption{Final Zy, Dy, and Th abundances as a function of $Y_e$ for the disk wind using the FRDM2012 mass model.\label{fig:ye}}
\end{figure}

Figure~\ref{fig:ye} shows the final calculated Zr, Dy, and Th abundances as functions of $Y_e$.
At the highest values of $Y_e$ considered, a large amount of limited-\emph{r} material around the first peak (here Zr) is synthesized, yet material does not move much beyond the second \rp\ peak ($A\approx 130$, $Z\approx 54$) until $Y_e<0.30$.
With decreasing $Y_e$, the lanthanides (Dy) are produced, and actinide production begins at $Y_e<0.23$.
The oscillatory behavior of the lanthanide and actinide abundances at very low $Y_e$ are due to fission cycles that occur in very neutron-rich environments \citep[as discussed in detail in][]{holmbeck2018b}.

With final abundances generated as functions of $Y_e$, we randomly select fifteen $Y_e$'s between 0.005 and 0.450 and the corresponding final Zr, Dy, Th, and U abundances.
Next we add the total Zr, Dy, Th, and U abundances over the fifteen randomly selected values.
If the total \ZrDy, \ThDy, and $\log\epsilon({\rm U/Th})$ abundances are within the specified constraints of Table~\ref{tab:ratios}, we keep all fifteen $Y_e$'s.
We repeat this sampling until we accumulate 100 successes, summing
a total of 1500 individual abundance patterns.
When combined, the summed abundances pattern matches the relative observational Zr, Dy, Th, and U abundances for a given kinematic group within the listed tolerances.

% =====================================================
\section{ADM Model Results}
\label{sec:results}

\begin{figure}[t]
	\centering
    \includegraphics[width=0.9\columnwidth]{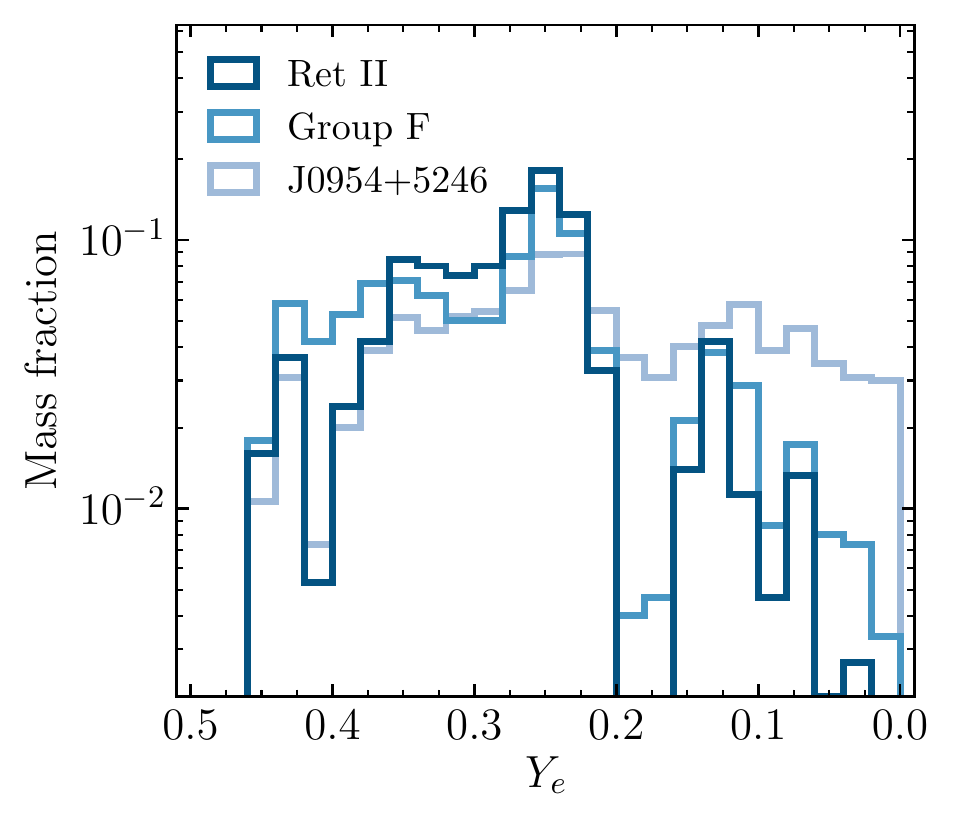}
    \caption{Ejecta distributions characterizing an \rp\ event predicted by the ADM model when matching Ret~II, Group~F, and J0954+5246 abundances using the disk wind trajectory and the FRDM2012 mass model.
    \label{fig:mass_all}}
\end{figure}

Figure~\ref{fig:mass_all} shows the empirical \rp\ ejecta distribution results of the ADM model applied to the three stellar cases discussed in Section~\ref{sec:kinematic_groups}.
The empirical mass ejecta distributions that characterize the observed abundance ratios of Ret~II, Group~F, and J0954+5246 mainly differ in the very low-$Y_e$ tail ($Y_e<0.18$ in this trajectory) where robust fission cycling and actinide production occurs.
The low actinide abundance constraints of the Ret~II group allows less mass in this very low-$Y_e$ tail to be ejected, while the actinide-normal Group~F and actinide-boost J0954+5246 allow increasing amounts of this fission-cycled material.

The bulk of the mass of material (at $Y_e\geq 0.18$) maintains a similar shape in all three cases, including a strong preference for $Y_e\approx 0.25$ and a dip in ejecta production at $Y_e\approx 0.18$.
The peak occurs because the \ThDy\ ratio is satisfied near $Y_e\approx 0.25$ for all three cases.
On the other hand, the dip at $Y_e\approx 0.18$ coincides with maximal actinide production and (locally) minimal lanthanide production when using this trajectory (see Figure~\ref{fig:ye}), producing a \ThDy\ ratio that is much higher than what observations suggest.

\begin{figure}[t]
	\centering
    \includegraphics[width=\columnwidth]{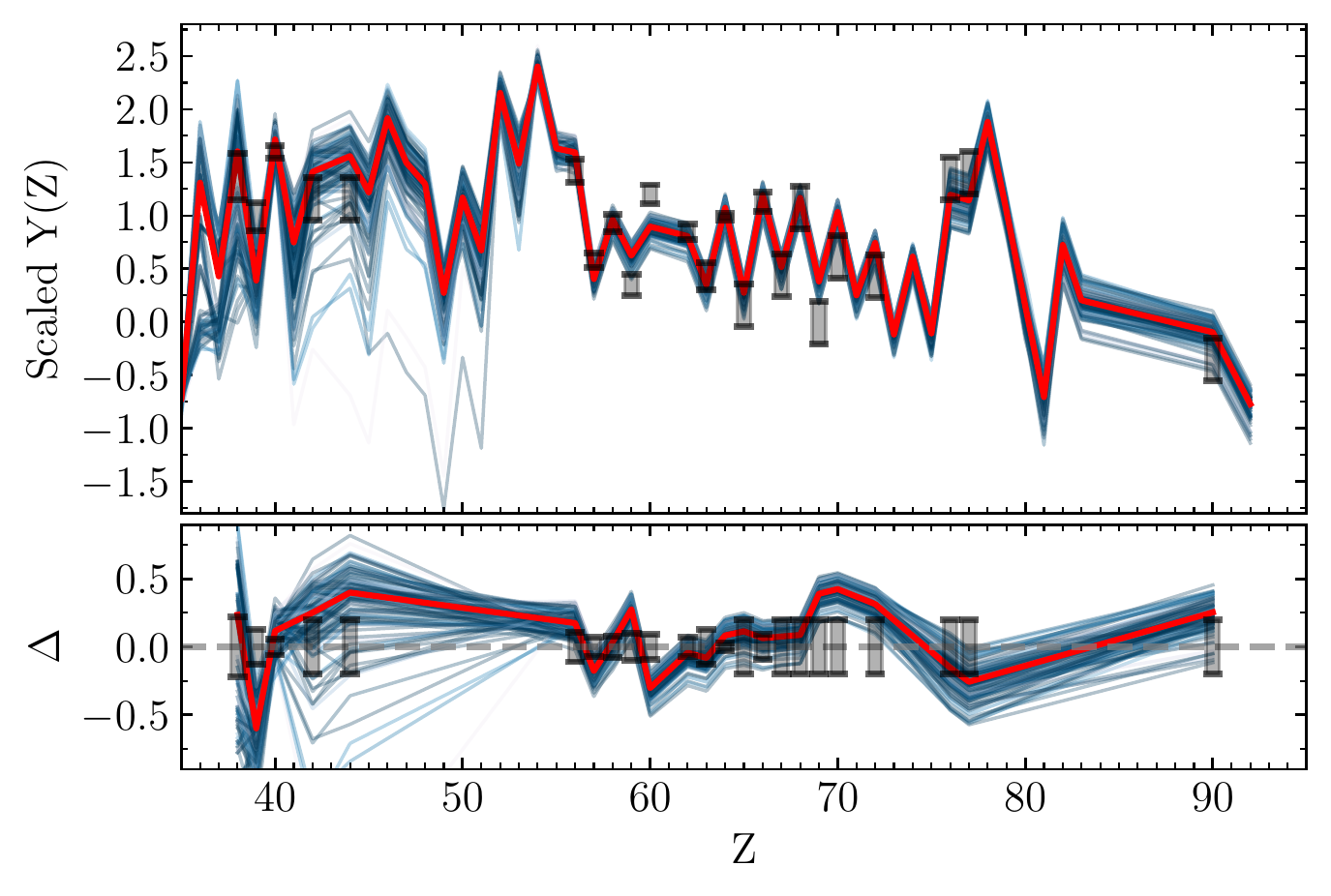}
    \includegraphics[width=\columnwidth]{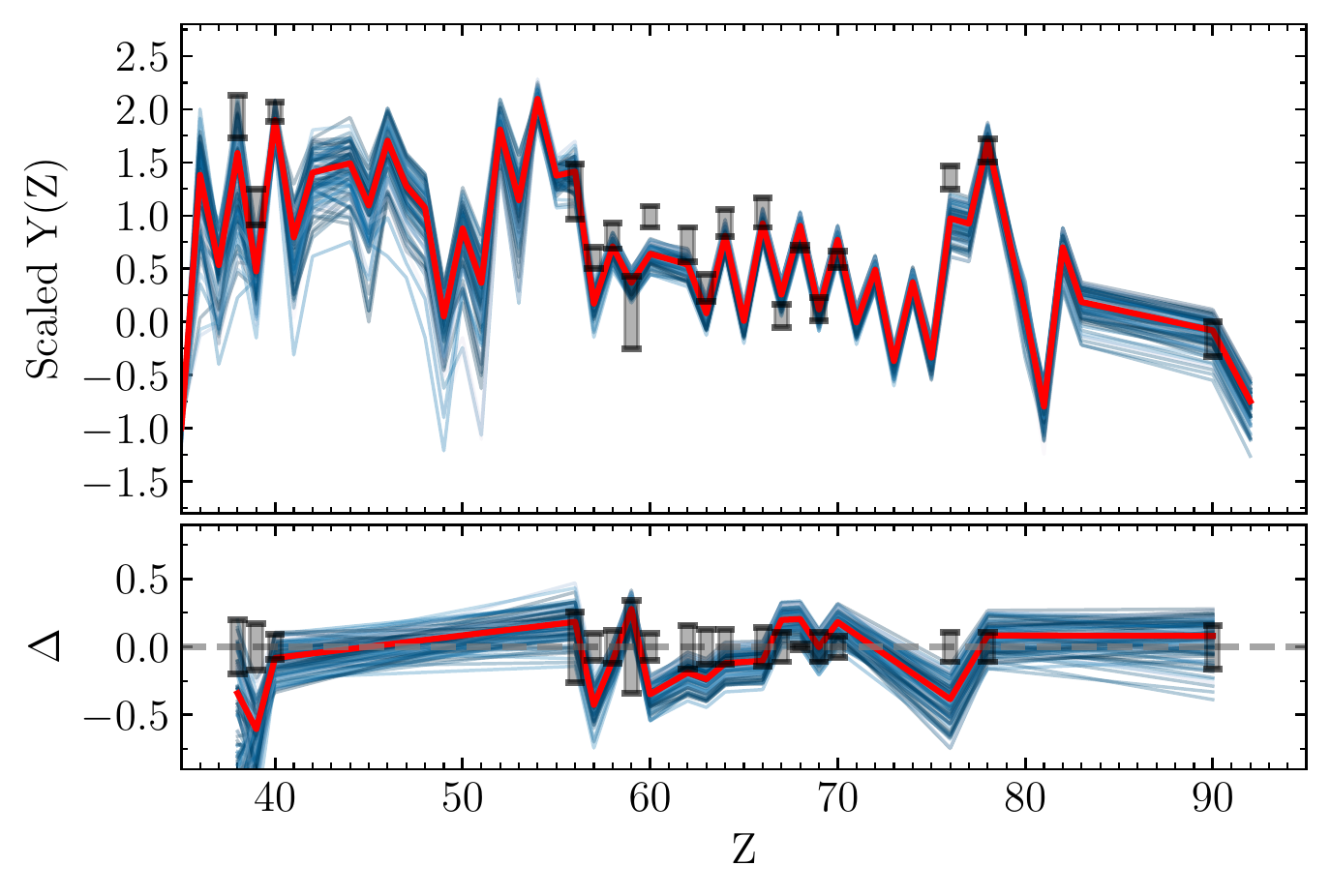}
    \includegraphics[width=\columnwidth]{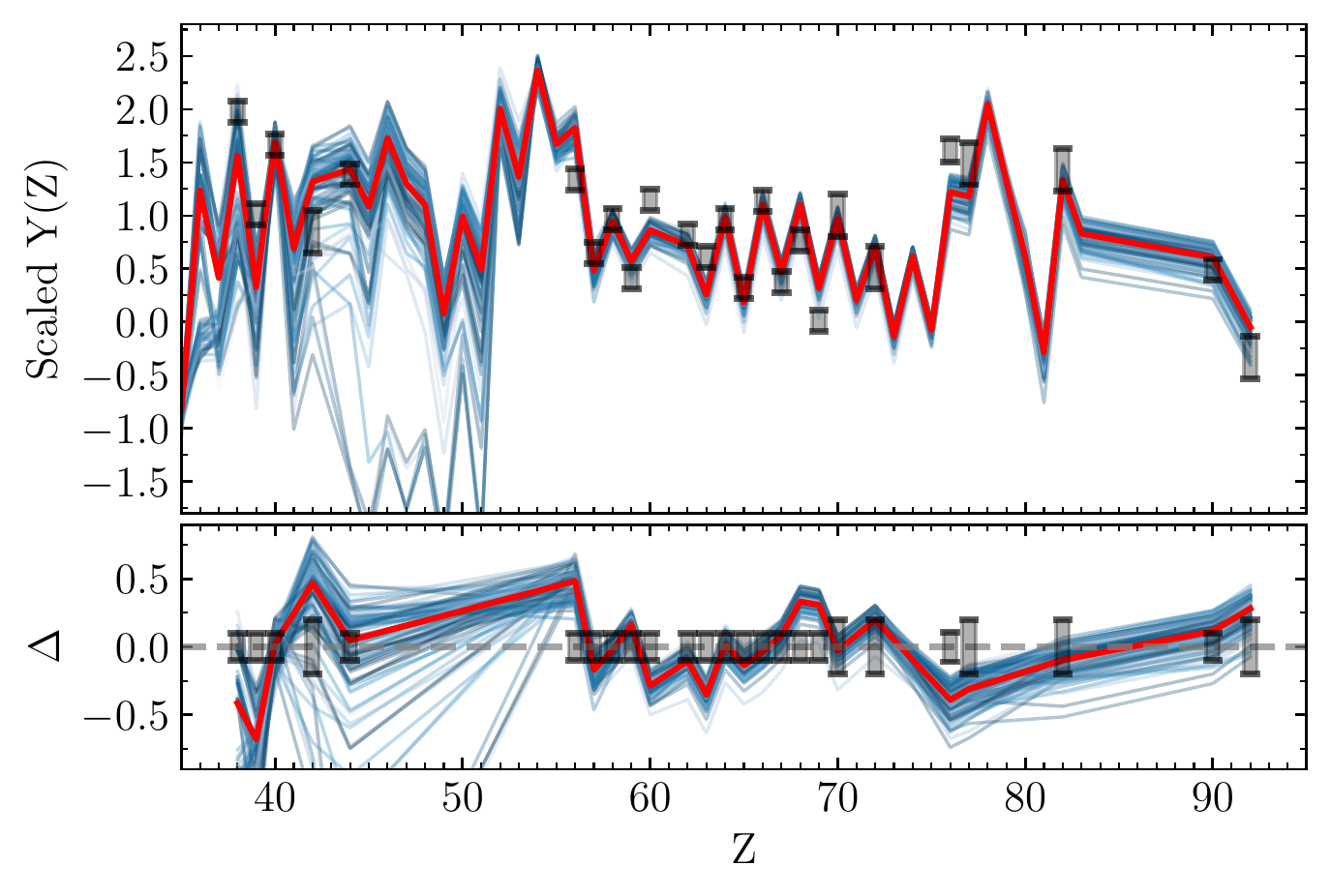}
    \caption{Final combined abundance pattern results (red lines) of the ADM model when matching Ret~II (top), Group~F (middle), and J0954+5246 (bottom) abundances. Successes of individual runs are shown in blue.\label{fig:ab_wind_frdm}}
\end{figure}

Figure~\ref{fig:ab_wind_frdm} shows the final abundance patterns for the ejecta distributions shown in Figure~\ref{fig:mass_all}.
Every individual abundance pattern (blue) represents a successful set of the fifteen random $Y_e$ choices made in the ADM method.
Each combined abundance pattern (red) succeeds in reproducing the scaled abundances of the limited-\emph{r} elements and many of the lanthanide elements.
The common dip surrounding $Z=60$ (Nd) is mostly due to the strong shell closures of FRDM2012, and partially due to the pure symmetric fission fragment yields we employ. However, this underproduction does not have any influence over the results we present here.
We finally note that for all three stellar groups, we have only supplied three abundance constraints to the ADM model. Hence, with few constraints, relatively good agreements across the entire \rp\ patterns are produced.

% =====================================================
\subsection{The Low-$Y_e$ Component}

\begin{figure}[t]
	\centering
    \includegraphics[width=0.95\columnwidth]{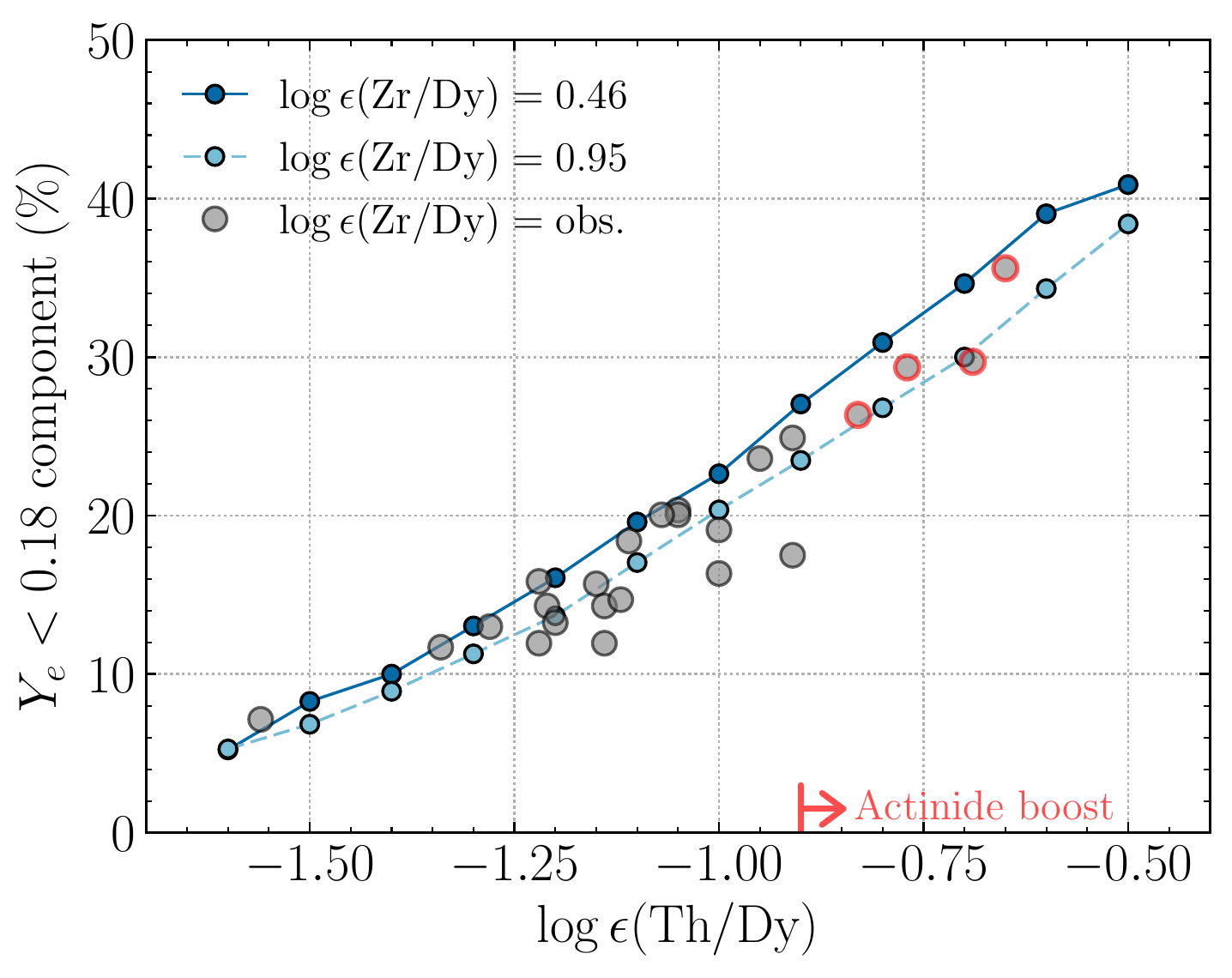}
    \caption{Percentage of allowed very low $Y_e$ ($<$0.18), actinide-rich mass to reproduce various \ThDy\ abundances, while requiring the specified \ZrDy\ ratio. Gray dots show the ADM model applied to select \rp\ stars with observed \ZrDy\ and \ThDy\ ratios as input constraints \citep{abohalima2017,placco2017,sakari2018,holmbeck2018}. \label{fig:lowYe_frdm2012}}
\end{figure}

The largest difference in the empirical $Y_e$ distributions of ejecta with varying levels of actinide enhancement lies in the allowed mass produced in very low-$Y_e$ environments.
To investigate this difference in detail, we systematically vary the ADM model input \ThDy\ constraint while holding the \ZrDy\ constraint constant. This way, we can quantify the amount of very low-$Y_e$ material that the progenitor \rp\ event may eject.
We repeat this process twice, once holding the \ZrDy\ constraint at 0.46 and again at 0.95, following the labeled bounds in Figure~\ref{fig:scatter} (top panel).
Recall that these bounds contain \rp\ enhanced stars in which the \rp\ material likely originated from one \rp\ event.
These systematic results are also compared to ADM results using both the \ThDy\ and \ZrDy\ observational abundance ratios from single \rp\ enhanced stars in the bottom panel of Figure~\ref{fig:scatter}.

Systematically varying the \ThDy\ input constraint shows a smoothly increasing fraction of allowed ejecta masses at very low-$Y_e$.
The \rp\ enhanced stars with likely single \rp\ progenitors fall between the two calculated curves (blue solid and dashed lines) in Figure~\ref{fig:lowYe_frdm2012}, by definition.
Most of these stars thus allow about 10\% to 25\% of their progenitor's \rp\ ejecta mass to be at $Y_e<0.18$.
The actinide-boost stars found at $\mbox{\ThDy}>-0.90$ allow roughly 25\% to 35\% of this very low-$Y_e$ material.
This enhancement accounts for increased actinide abundances.
Stars falling below the lower curve are those with higher \ZrDy\ ratios, which formed from gas that was likely polluted by multiple events.
Assuming the \rp\ signature in stars with higher \ZrDy\ originated from a single event, the ADM model can then account for their observed \rp\ element distributions using a mass ejecta distribution that is shifted to higher-$Y_e$ values.

Our ADM model results do not indicate a clear separation between the actinide-boost stars and their non-actinide-enhanced counterparts.
This agrees with the observed actinide abundances which suggest a smooth distribution of actinide enhancements, with the actinide-boost stars populating a low-probability tail of this distribution.
This indicates that the same \rp\ source can produce all levels of actinide enrichment seen in \rp\ enhanced stars. Different levels of actinide enhancement would then reflect a slightly different distribution in the mass ejecta properties within the \rp\ progenitor.
In all cases, the amount of fission-cycled (in this trajectory, $Y_e<0.18$) material required to reproduce our \rp\ abundance observations may be a significant---but not dominant---fraction of the entire \rp\ mass ejecta since it sensitively affects the actinide contribution.

\begin{figure}[t]
	\centering
    \includegraphics[width=0.9\columnwidth]{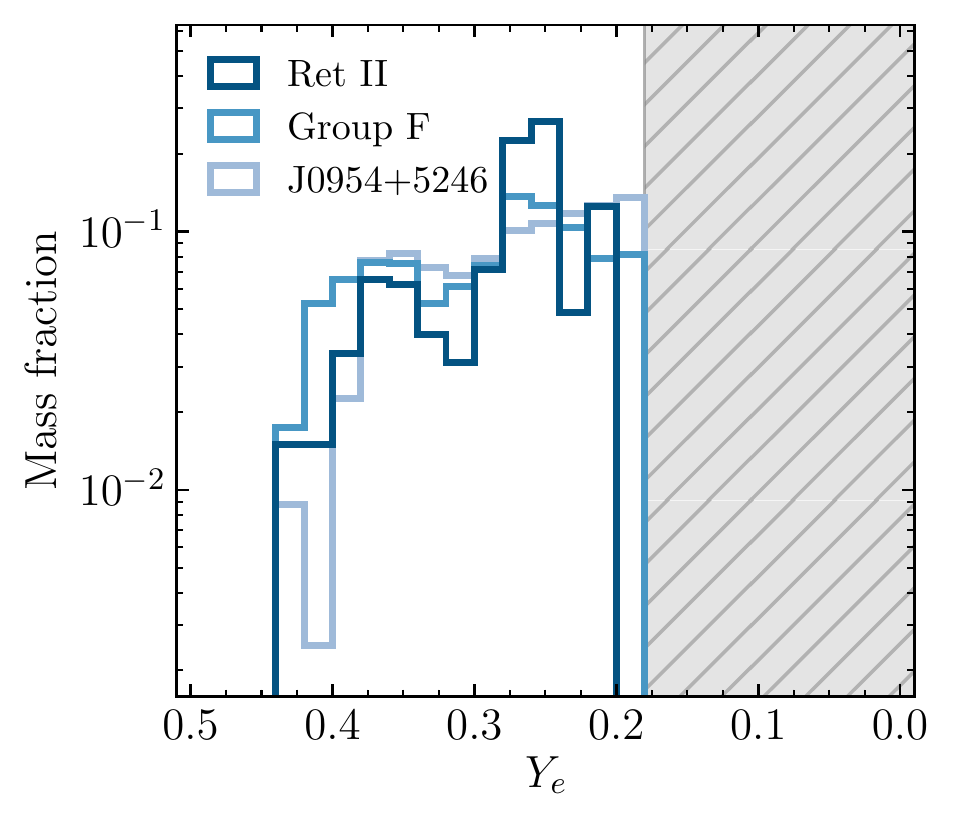}
    \caption{Ejecta distribution results for all three cases allowing no $Y_e<0.18$.\label{fig:nolow}}
\end{figure}

Interestingly, the abundance ratios can still be reproduced by the ADM model when the very low-$Y_e$ component is omitted entirely.
%SWITCHED THE INEQUALITIES
We investigate this effect by repeating the ADM calculation, only allowing the model to sample at $0.18\leq Y_e\leq 0.45$.
These results are shown in Figure~\ref{fig:nolow}.
Disallowing $Y_e$ below 0.18 produces a somewhat bimodal distribution driven by the \ZrDy\ and \ThDy\ requirements.
For Ret~II and Group~F, a peak forms at $Y_e\approx 0.25$, coinciding with the single $Y_e$ that satisfies the input \ThDy\ ratio.
%ADDED A WORD
Since now no Th can come from $Y_e<0.18$, all the Th contribution is concentrated around this $Y_e$.
However, for the actinide-boost case, not enough Th is produced at $Y_e\approx 0.25$, and the ejecta mass builds up near the cutoff at $Y_e=0.18$ where actinides are still able to be synthesized at levels necessary to eventually reproduce observed stellar abundances, within the allowed ranges of Table~\ref{tab:ratios}.
With the total amount of Dy constrained mostly by contributions from the $Y_e=0.25$ region, the Zr abundance primarily comes from higher values of $Y_e$.
This restraint produces the broad peak around $Y_e=0.37$.
Although these precise $Y_e$ constraints are mildly dependent on other astrophysical parameters (discussed in Section~\ref{sec:variations}), we conclude that it is possible to reproduce the abundance patterns seen in \rp\ enhanced stars without fission cycling (for the conditions considered here, meaning without $Y_e < 0.23$ material), but such a cutoff places stricter and more finely tuned requirements on the distribution of $Y_e$ in the ejecta.

The ADM model would fail for Ret~II if a $Y_e$ cutoff of 0.23 or greater was applied because there is simply not enough actinide material produced.
Similarly, applying a cutoff at $Y_e\geq 0.21$
would prevent the ADM model from reproducing actinide-boost abundance ratios.
As seen in Figure~\ref{fig:ye}, the Th abundance rises rapidly as $Y_e$ decreases from 0.24 to 0.17, covering over four dex---and thus all observed levels---of actinide abundance.
It is therefore unsurprising that the ADM model consistently favors this range.
Next, we turn to the higher $Y_e$ component which contributes the bulk of the ejected Zr (i.e., limited-\emph{r}) material.

% =====================================================
\subsection{The Higher-\emph{$Y_e$} Component}

\begin{figure}[t]
	\centering
    \includegraphics[width=0.95\columnwidth]{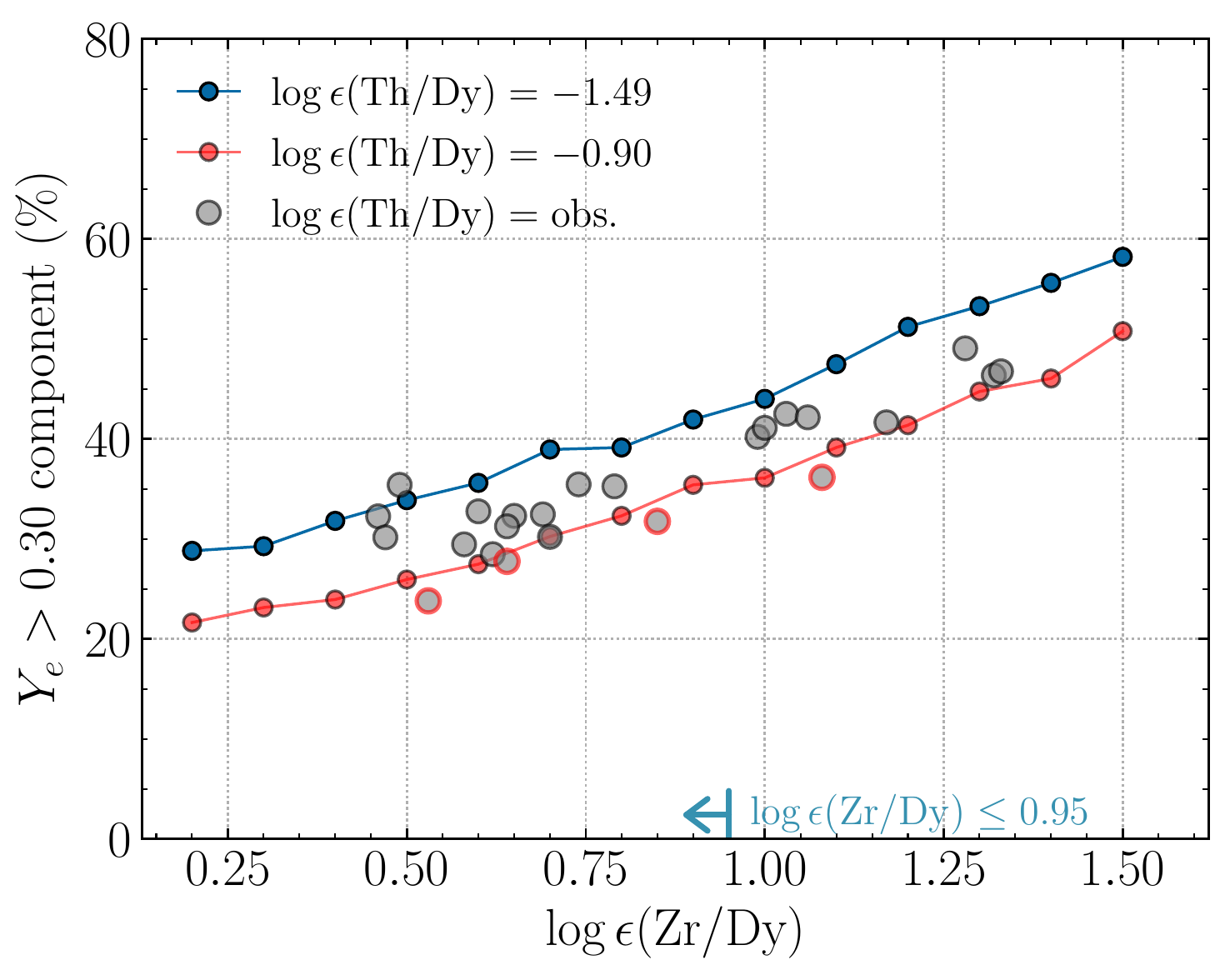}
    \caption{Percentage of allowed $Y_e>0.3$, Zr-rich mass to reproduce various \ZrDy\ abundances with constant \ThDy\ ratio constraints. Gray dots show the ADM model applied to select \rp\ stars with their observed \ZrDy\ and \ThDy\ ratios as input constraints \citep{abohalima2017,placco2017,sakari2018,holmbeck2018}, with red circles denoting the actinide-boost stars.
    \label{fig:highYe}}
\end{figure}

In analogy to Figure~\ref{fig:lowYe_frdm2012} of the very low-$Y_e$ component fraction, Figure~\ref{fig:highYe} shows the allowed fraction of material ejected at $Y_e>0.30$ as a function of the input \ZrDy\ constraint to characterize the limited-\emph{r} contribution from single \rp\ events.
The ADM model is run multiple times varying the input \ZrDy\ while holding the \ThDy\ constant, first at the actinide-boost cutoff ($-0.90$) and then at the very actinide-poor value following Ret~II ($-1.49$).
The ejecta mass fraction with $Y_e>0.30$ is also shown for individual stars using their observational \ZrDy\ and \ThDy\ abundance ratios as constraints.

Figure~\ref{fig:highYe} suggests that in order for the \rp\ event to synthesize all the required limited-\emph{r} material as well as the main \rp\ material, a minimum of roughly 25\% of the mass must be ejected at $0.30<Y_e \leq 0.45$.
For stars with $\mbox{\ZrDy} \leq 0.95$---which likely received their \rp\ material from only one progenitor---between roughly 25\% and 35\% of the progenitor ejecta mass has $0.30<Y_e \leq 0.45$.
Furthermore, because there is an observational minimum of $\mbox{\ZrDy}\approx 0.46$, our ADM model results imply that at least $\sim$25\% of the \rp\ ejecta mass must be ejected at these higher values of $Y_e$.

If the material in stars with $\mbox{\ZrDy} > 0.95$ were to originate from a single \rp\ progenitor, then more than 40\% of the \rp\ ejecta must be at $Y_e > 0.30$.
However, as previously noted, the main \rp\ material found in stars moderately enhanced in \rp\ elements with $\mbox{\ZrDy} > 0.95$ could have been diluted by limited-\emph{r}-only events such as CCSN neutrino-driven winds that primarily produce the limited \rp\ elements \citep{arcones2013,wanajo2013}.

% =====================================================
\section{Model Variations}
\label{sec:variations}

In this section, we investigate the impact that both astrophysical and nuclear physics variations have on the results of our ADM model to test the robustness of these empirically built mass ejecta distributions.

\subsection{Astrophysical Sites}

\label{sec:sites}

\begin{figure}[t]
	\centering
    \includegraphics[width=0.9\columnwidth]{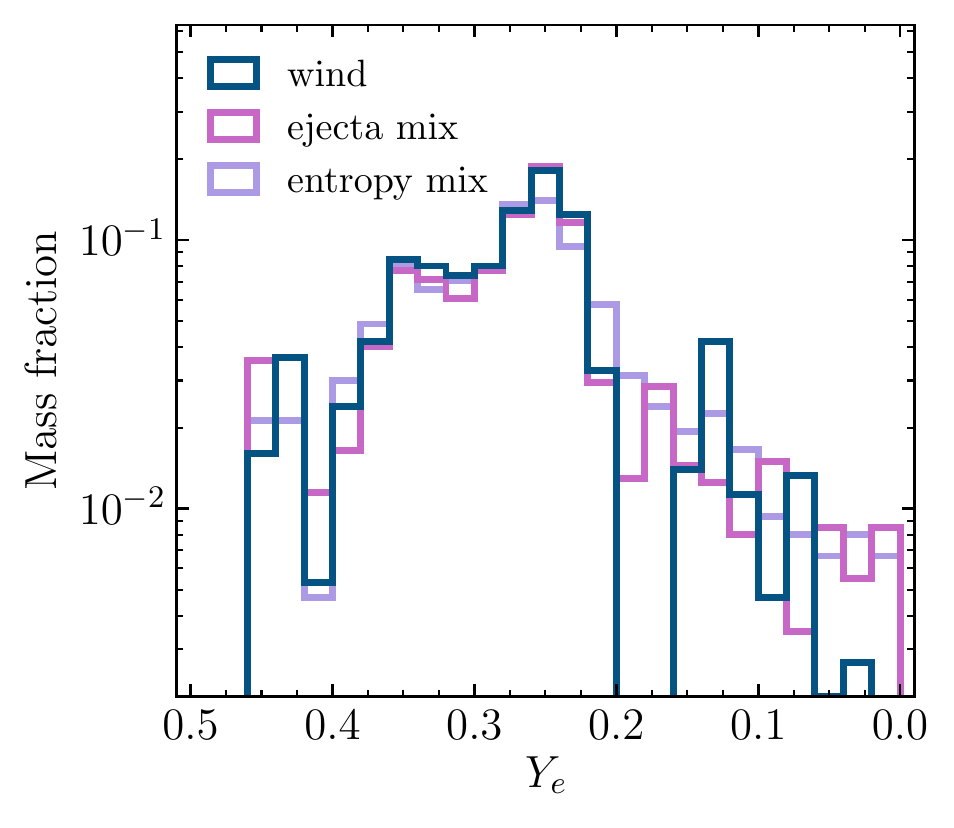}
    \caption{Ejecta distribution predicted by the ADM model matching Ret~II abundances using the disk wind trajectory only (``wind"), an NSM represented by a combination of wind and tidal ejecta (``ejecta mix"), and a combination of different entropies (``entropy mix"). All simulations use the FRDM2012 mass model.\label{fig:mass_mix}}
\end{figure}

The previous calculations only consider the \rp\ originating from a single site: an accretion disk wind.
Two situations that might occur in ``realistic" astrophysical \rp\ events are a mix of ejecta types and a mix of different entropies.
One promising \rp\ production site is the very low-$Y_e$ tidal ejecta of an NSM.
We choose a low-entropy ($s/k\approx 10$) trajectory from the 1.4--1.4 M$_\sun$ NSM simulations by S.\ Rosswog as in \citet{korobkin2012} for the tidal ejecta.
Next, we vary the initial $Y_e$ between 0.005 and 0.180 and run full \rp\ calculations for this tidal ejecta trajectory.
Then we used the ADM model to randomly sample from only the tidal ejecta component at $Y_e < 0.13$, and from only the wind component at $Y_e\geq 0.18$.
For the region at $0.13 \leq Y_e < 0.18$, the ADM model samples from both the tidal and wind ejecta with equal probability, producing a mixed-ejecta distribution.
This combination may be one representation of total NSM ejecta undergoing an \rp.
Figure~\ref{fig:mass_mix} shows the empirical $Y_e$ distribution obtained by using a combination of wind and tidal ejecta which match the Ret~II abundances (``ejecta mix").
Although the $Y_e<0.18$ component is distributed differently in the mixed ejecta case than the wind-only counterpart, the amount of necessary $Y_e<0.18$ mass from the tidal ejecta is similar to that of the wind.

The \rp\ can also feasibly occur in an environment that supports a range of entropies.
We investigate the effect of entropy on the $Y_e$ distribution by repeating the simulations with a high entropy ($s/k\approx 85$) trajectory for the entire range of $0.005 \leq Y_e \leq 0.450$ in equal steps, and extended the very low entropy tidal trajectory to $Y_e \leq 0.250$.
Next, the ADM model was run, randomly sampling between the original disk wind trajectory and the high entropy trajectory for $0.250 < Y_e \leq 0.450$, and between the low, medium, and high entropy trajectories for $0.005 \leq Y_e \leq 0.250$.
The ejecta distribution results from the ADM model using a random combination of entropies are shown in Figure~\ref{fig:mass_mix} (``entropy mix").

The previously mentioned dip at $Y_e\approx 0.18$ disappears when combining trajectories with different astrophysical properties.
This is because the value $Y_e=0.18$ does not universally signify robust actinide production for all \rp\ trajectories.
In the lowest entropy (tidal ejecta) trajectory, the Th abundance peaks at the lower $Y_e$ of 0.125.
At $Y_e=0.18$, instead of a peak in Th production occurring---as that produced by the high and medium entropy (wind) trajectories---the very low-entropy tidal ejecta trajectory produces a Dy peak, allowing the mass at $Y_e\approx 0.18$ to satisfy the input abundance ratio constraints and wash out the apparent two-component $Y_e$ distribution.

In summary, considering variations in the astrophysical site slightly affects the details of the predicted ejecta mass distribution. However, qualitatively, the ADM model robustly suggests that if there is any low-$Y_e$ fission cycling ejecta component, it must be small compared to the \rp\ material ejected by the disk wind at higher $Y_e$.

% =====================================================
\subsection{Nuclear Physics Inputs}

\begin{figure}[t]
	\centering
    \includegraphics[width=0.9\columnwidth]{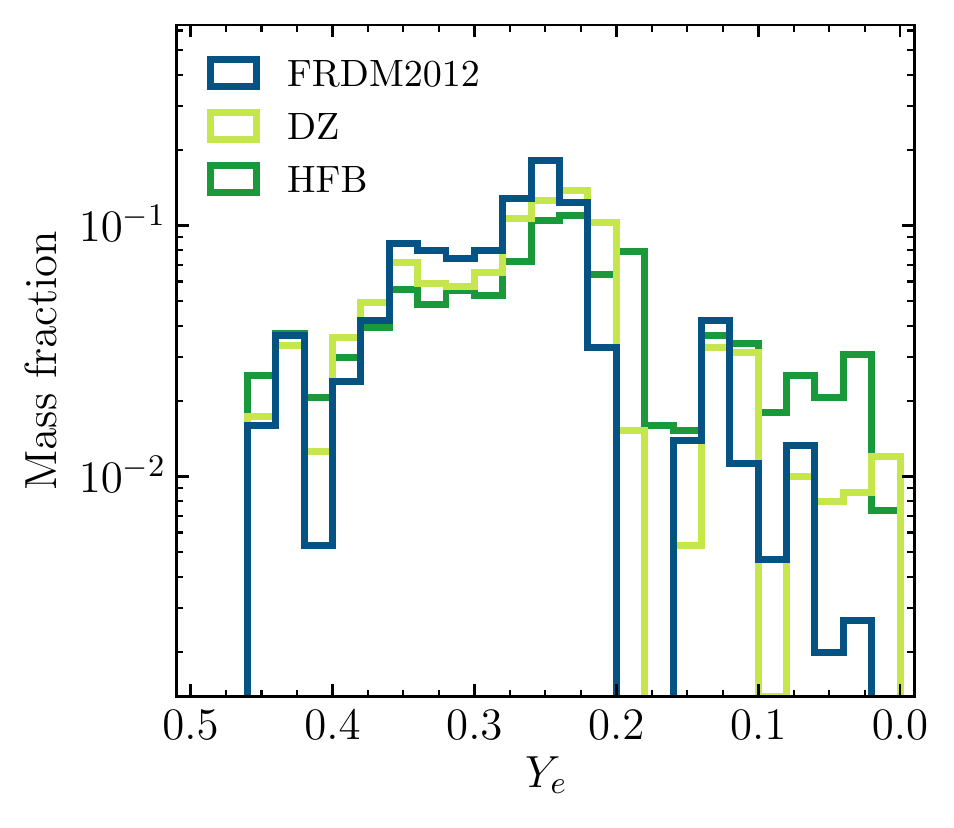}
    \caption{ADM predictions for Ret~II using a disk wind trajectory and the FRDM2012 (blue), DZ (light green), and HFB (dark green) mass models.\label{fig:mass_models}}
\end{figure}

Nucleosynthesis calculations of the \rp\ rely heavily on theoretical data to attempt estimates of reaction rates for very unstable (and as of yet unmeasured) nuclei along the \rp\ path.
Using different prescriptions of nuclear data far from stability can lead to dramatic differences in both the extent of the \rp\ and the final shape of the abundance pattern \citep[e.g.,][]{kratz1993,kratz1998,wanajo2004,mumpower2016}.
We test the robustness of the ADM model results by repeating our calculations using nuclear data informed by the Duflo-Zuker \citep[DZ;][]{duflo1995} and the Hartree-Fock-Bogoliubov \citep[HFB;][]{hfb17} mass models.
We use theoretical reaction and decay rates recalculated to be consistent with each mass model and use HFB barrier heights for fission rates within this mass model as in \citet{vassh2018}.
Figure~\ref{fig:mass_models} shows the results using the disk wind trajectory and three different nuclear mass models, using the Ret~II abundance constraints.
Although using DZ and HFB mass models results in ADM distributions with more low-$Y_e$ mass, the relative shape and magnitude of high-$Y_e$ material reflects our results found when using the FRDM2012 mass model.

% =====================================================
\subsection{The Low-$Y_e$ Component}

\begin{figure}[t]
	\centering
    \includegraphics[width=0.95\columnwidth]{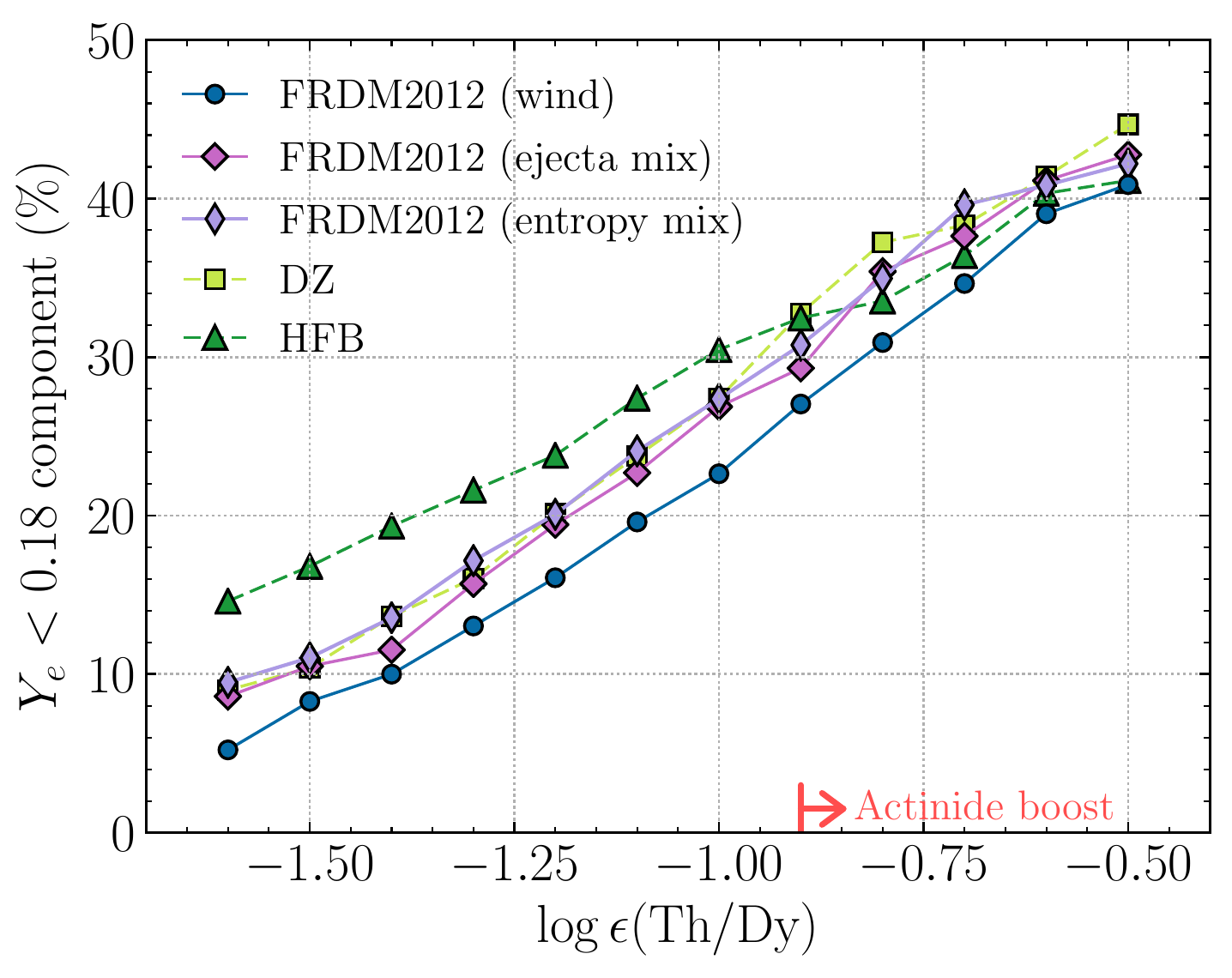}
    \caption{Percentage of allowed very low $Y_e$ ($<$0.18), actinide-rich mass to reproduce various \ThDy\ abundances constraining the \ZrDy\ to be the Ret~II value for each mass model specified. 
    \label{fig:lowYe_massmodel}}
\end{figure}

As seen in Figure~\ref{fig:mass_models}, the amount of predicted low-$Y_e$ ejecta mass varies with mass model.
In contrast, Figure~\ref{fig:mass_mix} displays little variation when using a mix of ejecta types or entropies.
In Figure~\ref{fig:lowYe_massmodel}, we quantify the fraction of very low $Y_e$ mass that the ADM model predicts is ejected when applying nuclear and astrophysical variations across a range of actinide abundances.
The DZ mass model tends to allow $\sim$5\% more very low $Y_e$ material than FRDM2012 since simulations using the DZ mass model does not produce the actinides as robustly as with FRDM2012 \citep{holmbeck2018b}.
Similarly, material leaves the actinide region due to higher neutron-induced reaction flows at later times with the HFB mass model compared to when using FRDM2012, also producing a lower final actinide abundance.
As a result, using HFB masses allows for $\leq$10\% more low-$Y_e$ mass than when employing FRDM2012.
Using a combination of tidal and wind ejecta or a combination of entropies slightly boosts the allowed very low $Y_e$ mass; however, the astrophysical variations lie comfortably within uncertainties in the nuclear masses.

In summary, accounting for nuclear mass model variations, the very low $Y_e$ ejected mass fractions may be as high as 40\% to account for most observations of actinides in \rp\ enhanced metal-poor stars.
Our results are robust under changes to the nuclear physics, with a variation of the allowed low-$Y_e$ component of $\sim$10\% of the total mass when considering variations to nuclear mass models or astrophysical environments.

% =====================================================
\section{The GW170817 Associated Kilonova}
\label{sec:gw710817}

A parameterized accretion disk trajectory---and the conclusions drawn from using this trajectory---is consistent with one possible description of NSM ejecta environments.
However, these conclusions are not necessarily unique and could still be applicable for other astrophysical sites, such as collapsars.
Here we test if the ADM model results agree with what has been inferred from the GW170817 associated kilonova (``SSS17a" or ``AT\,2017gfo"). This could offer another hint for NSMs as primary \rp\ sources of material in early small, emerging dwarf galaxies that gave rise to the \rp\ enhanced stars.

\citet{cowperthwaite2017} proposed that the light curve AT\,2017gfo could only be explained by multiple components: a lanthanide-poor (``blue") and lanthanide-rich (``red") component.
Inspired by this two-component model, we split our ADM model $Y_e$ distributions for Ret~II into a blue and red component.
We define the blue component as primarily producing limited-\emph{r} elements, extending over a range of $0.29<Y_e\leq 0.45$ (recall Figure~\ref{fig:ye}), and the red component consisting of the remaining material at $Y_e\geq 0.29$.
Using these ranges, we find lanthanide mass fractions of $X_{\rm lan}=10^{-0.8}$ for the red and $X_{\rm lan}=10^{-3.8}$ for the blue components when using our ADM model.
The mass ratio between these components is $m_{\rm red}/m_{\rm blue} = 1.7$.

\citet{kasen2017} also invoke a two-component model to resemble AT\,2017gfo, based on the high opacity of lanthanide elements that would produce an extended emission spectrum.
For their models to agree with AT\,2017gfo, a lanthanide-rich red kilonova would need to have a lanthanide mass fraction of $X_{\rm lan}\sim 10^{-1.5}$, and the lanthanide-poor blue kilonova would need to have $X_{\rm lan}\sim 10^{-4}$.
The ejecta mass ratio they estimate between these components is $m_{\rm red}/m_{\rm blue} = 1.6$.

The lanthanide mass fractions extracted from our model are slightly larger than those found by \citet{kasen2017}.
Our ADM simulations do not extend to iron-peak elements, which could be produced in higher-$Y_e$ regions during a NSM event.
Adding a contribution from iron-peak ejecta could bring our lanthanide mass fractions into further agreement with results by \citet{kasen2017}.
Overall, our results agree, despite our inherently different approaches.

% =====================================================
\section{Conclusions}

Using elemental abundances of \rp\ enhanced metal-poor stars, we have constructed empirical $Y_e$ distributions describing the ejecta of \rp\ events through the ADM model.
We find that the \rp\ abundance signatures of actinide-boost and actinide-deficient stars can likely originate from variations in $Y_e$ distribution of ejecta from the same type of astrophysical \rp\ event.
Both observationally and in the ADM model results, there is no clear point or distinct set of conditions at which the actinide-boost activates.
Rather, the smoothness of the distribution of observed actinide abundances correlates well with the smooth growth of the allowed very low $Y_e$ tail of our ADM ejecta mass distributions, as seen in Figures~\ref{fig:lowYe_frdm2012} and \ref{fig:lowYe_massmodel}.

Most actinide enrichments of metal-poor \rp\ enhanced stars can be explained by an \rp\ source with a very neutron-rich fission-cycling component. We estimate this fission-cycling ejecta to be a non-dominant (10--30\%) constituent of the \rp\ ejecta mass.
All levels of limited-\emph{r} abundance with respect to the lanthanides in stars with $\mbox{\ZrDy}\leq 0.95$ can be straightforwardly accommodated within the same \rp\ source.
For these stars, the lanthanide-poor component but which is rich in limited-\emph{r} elements, constitutes about 25--40\% of the ejecta mass.
This suggests that the \rp\ material in these stars need only come from one site that can produce the entire observed relative \rp\ abundance range from Sr to U.

The \rp\ signatures of very metal-poor stars allow the study of single \rp\ events, which we have characterized through the ADM model.
We compared our empirically found progenitor $Y_e$ distributions of ejecta to the results of an independent study of the currently favored \rp\ site, an NSM.
We found that both the lanthanide mass fraction and the red-to-blue mass ejecta ratio derived from the ADM model are consistent with results matching the light curve of the GW170817 associated kilonova, AT\,2017gfo.
The shape of our empirical $Y_e$ distributions also resemble those extracted from available hydrodynamical NSM simulations \citep{fernandez2015,radice2018}.
However, the accretion disk wind used in this work may be theoretically similar to---or perhaps even observationally indistinguishable from---other astrophysical sites, e.g., the accretion disk wind from a collapsar remnant.
Future LIGO/aLIGO detections of NSMs and follow-up observations of their electromagnetic counterparts will be helpful to further characterize the progenitor site(s) of \rp\ enhanced stars.

In addition to investigations of NSMs and other \rp\ events, a comprehensive study of the \rp\ calls for more observations of metal-poor stars enhanced in these elements.
Further identifications of \rii\ stars and their elemental abundances can be used to progress several areas of \rp\ studies. 
For example, more measurements of Th can test if DES~J033523$-$540407 in Ret~II and J0954$+$5246 in the halo represent limits on Th/Dy production, or if an even broader range exists.
Large actinide variations at higher metallicities could indicate activity by other \rp\ sources as a function of chemical evolution, which can be identified and characterized through theoretical tools such as ADM.
Detailed spectroscopy of more \rii\ stars will also allow further measurements of U.
Due to the observed spread in abundance ratios, there is currently no unifying set of actinide-to-lanthanide production ratios that can be unilaterally applied to carry out cosmochronometry.
However, the U/Th ratio principally remains a robust and reliable tool for radioactive decay dating if ejecta distributions built from observed element patterns could be used to refine the required type of production ratios that accurately reflect the relevant progenitor site(s).

Overall, larger numbers of known \rii\ stars would increase e.g., identifications of kinematic groups in the Galactic halo or enable additional Th and U measurements. Hence, a main objective of the \emph{R}-Process Alliance \citep[RPA;][]{hansen2018,sakari2018,aprahamian2018} is to increase the number of known \rii\ stars from $\sim$30 to $\sim$100. Applying information from additional statistically significant kinematic groupings to the ADM model could then be used to investigate whether NSMs are main sources of \rp\ material, or if the ADM suggests that other \rp\ sources are predominantly needed.

The wealth of stellar abundance data---from surveys such as that being conducted by the RPA---together with theoretical \rp\ studies, future NSM detections, and nuclear physics constraints from next-generation rare-isotope beam facilities (e.g., FRIB) will allow thorough investigations of the origins of all \rp\ elements.

\acknowledgements
%The authors thank Trevor M.\ Sprouse and Matthew Mumpower for their extensive work in developing PRISM and the nuclear input used in this study.
This work was initiated by discussions at the FRIB Theory Alliance workshop, ``FRIB and the GW170817 kilonova" \citep{aprahamian2018}, which was supported by the U.S. Department of Energy (DOE) through the FRIB Theory Alliance under Contract No.\ DE-SC0013617.
The authors also benefited from support by the National Science Foundation (NSF) under Grant No.\ PHY-1430152 (JINA Center for the Evolution of the Elements).
A.F.\ is partially supported by National Science Foundation grant AST-1716251.
This work was supported in part by the U.S.\ DOE under contracts Nos.\ DE-FG02-02ER41216 (G.C.M.), DE-FG02-95-ER40934 and DE-SC0018232 (R.S.\ and T.M.S.), and DE-AC52-07NA27344 for the topical collaboration Fission In R-process Elements (FIRE; G.C.M., R.S., and M.R.M.), and by the NSF under contract No.\ PHY-1630782 (N3AS; G.C.M. and R.S.).
M.R.M.\ was supported by the US DOE through the Los Alamos National Laboratory operated by Triad National Security, LLC, for the National Nuclear Security Administration of U.S. DOE (Contract No.\ 89233218CNA000001).

%\software{Python}

% =====================================================
\bibliography{bibliography}

% =====================================================

\end{document}